\documentstyle[aps,twocolumn,epsf,psfig]{revtex}
\begin{document}
\title{On the fulfillment of Ward identities in the functional renormalization
group approach }
\author{A. A. Katanin$^{a,b}$}
\address{$^a$ Max-Planck-Institut f\"ur Festk\"orperforschung, D-70569, Stuttgart\\
$^b$ Institute of Metal Physics, 620219 Ekaterinburg, Russia}
\maketitle

\begin{abstract}
I consider the fulfillment of conservation laws and Ward identities in the
one- and two-loop functional renormalization group approach. It is shown
that in a one-particle irreducible scheme of this approach Ward identities
are fulfilled only with the accuracy of the neglected two-loop terms $%
O(V_\Lambda ^3)$ at one-loop order, and with the accuracy $O(V_\Lambda ^4)$
at two-loop order ($V_\Lambda $ is the effective interaction vertex at scale 
$\Lambda $). The one-particle self-consistent version of the two-loop RG
equations which leads to smaller errors in Ward identities due to the
absence of the terms with non-overlapping loops, is proposed. In particular,
these modified equations exactly satisfy Ward identities in the ladder
approximation.

\vspace{0.4cm}
\end{abstract}

Strongly-correlated systems attract currently a lot of interest. While
rigorous mathematical methods, such as the Bethe ansatz exist in one
dimension, the consideration of systems with the dimensionality $d>1$
requires certain approximations. These approximations can break conservation
laws and therefore lead to results which are in contradiction with the
conservation of particle number, momentum, energy, etc.

The fulfillment of Ward identities \cite{Ward,Abrikosov,Toyoda} guarantees
the conservation of different quantities. Many years ago a way to construct
conserving approximations was proposed by Baym and Kadanoff \cite
{BaymKadanoff}. They have shown that the sufficient condition to fulfill
conservation laws is the $\Phi $-derivability, which requires that for a
given generating functional $\Phi [G]$ the self-energy $\Sigma $ and the
two-particle vertex $\overline{\Gamma }$ irreducible in the particle-hole
channel satisfy $\Sigma =\delta \Phi /\delta G$ and $\overline{\Gamma }%
=\delta ^2\Phi /\delta G^2$.

Later on, motivated by these ideas, Bickers and Scalapino \cite{FLEX}
proposed the fluctuation exchange approximation (FLEX), which is based on a
certain choice of $\Phi [G]$. This approximation has the advantage that it
satisfies the conservation laws by construction. However, the disadvantage
of FLEX is that the vertex $\overline{\Gamma }=\delta \Sigma /\delta G$
which satisfies Ward identities is obtained only on the next level of
approximation when the self-energy is already computed and therefore the
calculation of the self-energy is performed with the vertices which do not
satisfy Ward identities (see e.g. the discussion in Ref.\cite{Tremblay}).

The situation is somewhat better in parquet-type approaches \cite{Parq} (see
also Refs. \cite{Bickers1,JanisR}). Both, vertex- and self-energy effects
are accounted for in these approaches on the same ground. Although general $%
\Phi $-derivability of the parquet approach was recently proven in Ref. \cite
{Janis}, the practical application of these approaches for systems with $d>1$
meets serious computational difficulties and was performed only in a
restricted number of cases \cite{Janis1,DzyJak}.

Recently proposed functional renormalization group (RG) techniques, based on
Polchinskii equations \cite{Polchinskii,Zanchi}, Wick-ordered RG equations 
\cite{SalmBook,Metzner}, and one-particle irreducible (1PI) RG equations\cite
{SalmHonR,SalmHon,SalmHon1}, consider the self-energy and vertex corrections
on the same footing, but require much smaller computational effort. These
techniques were recently applied to the calculation of two-particle \cite
{Zanchi,SalmHon,SalmHon1,SalmHon2,KK} and one-particle \cite{SalmHonSE,KK1}
properties of the two-dimensional Hubbard model, and to the impurity problem
in a Luttinger liquid \cite{MetzAnd}.

The fulfillment of Ward identities within the functional RG techniques is a
non-trivial problem because of the use of different approximations: the
truncation of an infinite hierarchy of RG equations (loop expansion),
projection of the vertices on the Fermi surface etc. The main restriction
comes from the loop expansion, used within these approaches. Unlike previous
field-theoretical and Wilson RG approaches, not only singular, but also
regular terms are accounted for in functional RG. This can lead to some
inconsistencies and to the violation of Ward identities. Besides the
approximations mentioned above, the introduction of the cutoffs in momentum
or energy spaces can break Ward identities at the intermediate stages of the
RG flow\cite{Enss}.

Therefore, it is useful to verify whether Ward identities are fulfilled in
the functional RG approach. In the present paper I consider the fulfillment
of Ward identities in one- and two-loop 1PI functional RG analysis without
the projection of the vertices to the Fermi surface.

I consider a model with the action 
\begin{eqnarray}
{\cal S} &=&\sum_k(i\nu _n-\varepsilon _{{\bf k}})c_{k\sigma }^{\dagger
}c_{k\sigma }+\sum_{k_1...k_4,\sigma ,\sigma ^{\prime }}  \label{act} \\
&&V_{k_1k_2;k_3k_4}^0c_{k_1,\sigma }^{\dagger }c_{k_2,\sigma ^{\prime
}}^{\dagger }c_{k_3\sigma }c_{k_4\sigma ^{\prime }}\delta (k_1+k_2-k_3-k_4) 
\nonumber
\end{eqnarray}
with the general (frequency dependent) interaction $V_{k_1k_2;k_3k_4}^0$; $%
k_i=(i\nu _n^{(i)},{\bf k}_i)$.

The conservation of charge leads to a continuity equation, which can be
written in the continuum limit as 
\begin{eqnarray}
&&\ \ \ \ \ \ \ \ \ \ \ q_\mu J_{{\bf q}}^\mu 
\begin{array}{c}
=
\end{array}
0,  \label{cc} \\
J_{{\bf q}}^0 &=&\sum_{{\bf k},\sigma }c_{{\bf k},\sigma }^{\dagger }c_{{\bf %
k+q},\sigma };\,{\bf J}_{{\bf q}}=\sum_{{\bf k},\sigma }{\bf (}\nabla
\varepsilon _{{\bf k}}{\bf )}c_{{\bf k},\sigma }^{\dagger }c_{{\bf k+q}%
,\sigma }+{\bf J}_{{\bf q}}^{\text{int}},  \nonumber
\end{eqnarray}
where $\mu =0...3,$ $q=(\omega ,{\bf q}),$ ${\bf J}_{{\bf q}}^{\text{int}}$
is the contribution of electron-electron interaction to the current, ${\bf J}%
_{{\bf q}}^{\text{int}}={\bf 0}$ for a density-density interaction with $%
V_{k_1k_2;k_3k_4}^0=f(k_1-k_3).$ The Ward identity which follows from (\ref
{cc}) connects the one-particle self-energy $\Sigma _k$ with the charge- and
charge-current vertices, 
\begin{eqnarray}
&&\ \ \Gamma _{k,k^{\prime };k+q,k^{\prime }-q}^0 
\begin{array}{c}
=
\end{array}
V_{k,k^{\prime };k+q,k^{\prime }-q} 
\begin{array}{c}
=
\end{array}
\\
&&\ \ G_k^{-1}G_{k^{\prime }}^{-1}G_{k+q}^{-1}G_{k^{\prime }-q}^{-1}\langle
T[c_{k^{\prime }}^{\dagger }c_{k^{\prime }-q}c_k^{\dagger }c_{k+q}]\rangle _{%
\text{ir}}  \nonumber \\
&&\ \ \Gamma _{k,k^{\prime };k+q,k^{\prime }-q}^i 
\begin{array}{c}
=
\end{array}
\nonumber \\
&&\ \ (\nabla _i\varepsilon _{k^{\prime }})G_k^{-1}G_{k^{\prime
}}^{-1}G_{k+q}^{-1}G_{k^{\prime }-q}^{-1}\langle T[c_{k^{\prime }}^{\dagger
}c_{k^{\prime }-q}c_k^{\dagger }c_{k+q}]\rangle _{\text{ir}}+\Gamma _{\text{%
int}}^i  \nonumber
\end{eqnarray}
where index $ir$ stands for connected contributions, $G_k$ is full
one-particle Green function, $\Gamma _{\text{int}}^i$ is the contribution to
charge-current vertex produced by ${\bf J}_{{\bf q}}^{\text{int}}$. The
corresponding Ward identity for the model (\ref{act}) reads (see Refs. \cite
{Abrikosov,Toyoda,Toyoda1} for a derivation) 
\begin{eqnarray}
&&q_\mu \sum_{k^{\prime }}(-2\Gamma _{k,k^{\prime };k+q,k^{\prime }-q}^\mu
+\Gamma _{k,k^{\prime };k^{\prime }-q,k+q}^\mu )  \nonumber \\
&&\,\,\,\,\,\,\,\,\,\,\,\,\,\times G_{k^{\prime }}G_{k^{\prime }-q} 
\begin{array}{c}
=
\end{array}
\Sigma _k-\Sigma _{k+q},  \label{Ward}
\end{eqnarray}
where $\Gamma _\mu =(\Gamma _0,{\bf \Gamma )}$.

The functional renormalization-group approach leads to an infinite hierarchy
of differential equations for the self-energy $\Sigma $, two-particle vertex 
$V^{(4)}=V,$ and higher-order vertices $V^{(n)},$ $n>4.$ Truncated at the
one-loop order, i.e. after neglect of $V^{(n)}$ with $n>4$ this hierarchy
reduces to two integro-differential equations for the self-energy and the
two-particle vertex, which in 1PI RG scheme\cite{SalmHonR,SalmHon} have the
form 
\begin{mathletters}
\label{OneLoop}
\begin{eqnarray}
\frac{d\Sigma ^\Lambda }{d\Lambda } &=&V^\Lambda \circ S^\Lambda ,
\label{OneLoopA} \\
\frac{dV^\Lambda }{d\Lambda } &=&V^\Lambda \circ (G^\Lambda \circ S^\Lambda
+S^\Lambda \circ G^\Lambda )\circ V^\Lambda .  \label{OneLoopB}
\end{eqnarray}
In Eq. (\ref{OneLoop}) $\circ $ stands for the summation over intermediate
momentum-, frequency and spin-variables according to standard diagrammatic
rules, $\Lambda $ is the scaling parameter specified below. The initial
condition for Eqs. (\ref{OneLoop}) is $\Sigma ^{\Lambda _0}=0,$ $V^{\Lambda
_0}=V_0$ where $\Lambda _0\gg \max (|\varepsilon _k|).$ The propagators $S,G$
are connected by 
\end{mathletters}
\[
S^\Lambda =-(G^\Lambda )^2\frac d{d\Lambda }(G_0^\Lambda )^{-1},
\]
where $G_0^\Lambda $ is the non-interacting Green function at scale $\Lambda 
$. In particular, in the sharp momentum cutoff scheme \cite
{SalmHon,SalmHonSE} the propagators $G^\Lambda $ and $S^\Lambda $ are given
by 
\begin{equation}
\left\{ 
\begin{array}{c}
G_k^\Lambda  \\ 
S_k^\Lambda 
\end{array}
\right\} =\left\{ 
\begin{array}{c}
\theta (|\varepsilon _{{\bf k}}|-\Lambda ) \\ 
-\delta (|\varepsilon _{{\bf k}}|-\Lambda )
\end{array}
\right\} \frac 1{i\nu _n-\varepsilon _{{\bf k}}-\Sigma _k^\Lambda },
\label{GMC}
\end{equation}
in the sharp frequency cutoff scheme (at $T=0$) \cite{MetzAnd} by 
\begin{equation}
\left\{ 
\begin{array}{c}
G_k^\Lambda  \\ 
S_k^\Lambda 
\end{array}
\right\} =\left\{ 
\begin{array}{c}
\theta (|\nu |-\Lambda ) \\ 
-\delta (|\nu |-\Lambda )
\end{array}
\right\} \frac 1{i\nu _n-\varepsilon _{{\bf k}}-\Sigma _k^\Lambda },
\label{GFC}
\end{equation}
and in the temperature cutoff scheme \cite{SalmHon1} $\Lambda =T$ and 
\begin{equation}
G_k^T=\frac{T^{1/2}}{i\nu _n-\varepsilon _{{\bf k}}-\Sigma _k^T};\,\;S_k^T=-%
\frac 12\frac{(i\nu _n+\varepsilon _{{\bf k}})/T^{1/2}}{(i\nu _n-\varepsilon
_{{\bf k}}-\Sigma _k^T)^2}.  \label{GTC}
\end{equation}
Note that the results obtained in the present paper do not depend on the
sharpness of the cutoff procedure in momentum or energy space and are valid
for smooth cutoff functions as well.

Consider first the Ward identity (\ref{Ward}) for pure frequency shift $%
q=(\omega ,{\bf 0})$ within the momentum-cutoff RG approach (\ref{GMC}).
Written in differential form the corresponding identity reads 
\begin{equation}
\partial \Sigma _k/\partial k_0=\sum_{k^{\prime }}(V_{k,k^{\prime
};k^{\prime },k}-2V_{k,k^{\prime };k,k^{\prime }})G_{k^{\prime }}^2
\label{Ward1}
\end{equation}
where $k=(k_0,{\bf k}).$

To verify whether this identity is satisfied at each intermediate scale $%
\Lambda $ within the 1-loop RG scheme (\ref{OneLoop}), one can use the
following procedure: differentiate (\ref{Ward1}) with respect to $\Lambda $
and compare with RG equations (\ref{OneLoop}) differentiated with respect to 
$k_0.$ The following identities are helpful when performing this procedure: 
\begin{eqnarray}
\frac{\partial G_k^\Lambda }{\partial \Lambda } &=&S_k^\Lambda +(G_k^\Lambda
)^2\frac{\partial \Sigma _k^\Lambda }{\partial \Lambda } \\
\frac{\partial G_k^\Lambda }{\partial k_0} &=&-(G_k^\Lambda )^2(1-\frac{%
\partial \Sigma _k^\Lambda }{\partial k_0})  \nonumber \\
\frac{\partial S_k^\Lambda }{\partial k_0} &=&-2G_k^\Lambda S_k^\Lambda (1-%
\frac{\partial \Sigma _k^\Lambda }{\partial k_0})  \nonumber
\end{eqnarray}
In this way one finds that the identity (\ref{Ward1})\ is satisfied if and
only if 
\begin{eqnarray}
&&\frac{\partial V_{k,k+q;k+q,k}}{\partial k_0} 
\begin{array}{c}
=
\end{array}
2\sum\nolimits_pV_{k,k+p;k+p,k}V_{k+p,k+q;k+q,k+p}G_{k+p}^3  \nonumber \\
&&\ \
\,\,\,\,\,\,\,\,\,\;+2\sum%
\nolimits_p(V_{k,k+q+p;k+p,k+q}-V_{k,k+q+p;k+q,k+p})  \label{id} \\
&&\ \ \,\,\,\,\,\,\,\,\,\;\times
V_{k+p,k+q;k+p+q,k}(G_{k+q+p}^2G_{k+p}+G_{k+p}^2G_{k+p+q})  \nonumber \\
&&\ \
\,\,\,\,\,\,\,\,\,\,\,\,+2\sum%
\nolimits_pV_{k+q,k;k+p,k+q-p}V_{k+q-p,k+p;k+q,k}G_{k+p}^2G_{q-p+k} 
\nonumber \\
&&\frac{\partial V_{k,k+q,k;k+q}}{\partial k_0} 
\begin{array}{c}
=
\end{array}
2\sum\nolimits_pG_{k+p+q}^3[V_{k,k+p;k+p,k}V_{k+p,k+q;k+p,k+q}  \nonumber \\
&&\ \
\,\,\,\,\,\,\,\,\,\,\,%
\,+V_{k,k+p;k,k+p}(V_{k+p,k+q;k+q,k+p}-2V_{k+p,k+q;k+p,k+q})]  \nonumber \\
&&\ \ \ \
\,\,\,\,\,\,\,\,+\sum%
\nolimits_pV_{k,k+p+q;k+p,k+q}^2(G_{k+p+q}^2G_{k+p}+G_{k+p+q}G_{k+p}^2) 
\nonumber \\
&&\ \
\,\,\,\,\,\,\,\,\,\,\,+\sum%
\nolimits_pV_{k,k+q;k+p,q-p+k}^2(G_{k+p}^2G_{q-p+k}+G_{q-p+k}^2G_{k+p}) 
\nonumber
\end{eqnarray}
Note that these equations have the same structure as the RG equation for the
vertex, Eq. (\ref{OneLoopB}) with the replacements $dV/d\Lambda \rightarrow
\partial V/\partial k_0$ and $S\rightarrow G^2,$ and therefore can be
shortly written as 
\begin{equation}
\frac{\partial V^\Lambda }{\partial k_0}=V^\Lambda \circ [G^\Lambda \circ
(G^\Lambda )^2+(G^\Lambda )^2\circ G_\Lambda ]\circ V_\Lambda
\end{equation}
To see whether this equation is compatible with the RG equation for the
vertex, Eq. (\ref{OneLoopB}), one can differentiate Eq. (\ref{id}) once more
with respect to $\Lambda $ and compare it with (\ref{OneLoopB})
differentiated with respect to $k_0.$ After long algebraic manipulations one
obtains for the difference 
\begin{eqnarray*}
\frac \partial {\partial \Lambda }\left( \frac{\partial
V_{k,k+q;k+q,k}^\Lambda }{\partial k_0}\right) _{\text{Eq. (\ref{id})}}-%
\frac \partial {\partial k_0}\left( \frac{\partial V_{k,k+q;k+q,k}^\Lambda }{%
\partial \Lambda }\right) _{\text{Eq. (\ref{OneLoopB})}}  \label{VV}
\end{eqnarray*}
(and similar for $V_{k,k+q;k,k+q}^\Lambda $) two types of terms, which can
be written in a short notation as 
\begin{eqnarray}
&&V\circ G^2\circ (G^2\circ \frac{d\Sigma }{d\Lambda }-S\circ \frac{d\Sigma 
}{dk_0})\circ V  \nonumber \\
&&\ \ \ \ \ +G^\Lambda \circ V^\Lambda \circ G^\Lambda \circ V^\Lambda \circ
(G^\Lambda )^2\circ V^\Lambda \circ S^\Lambda .  \label{VVV}
\end{eqnarray}
Both terms in Eq. (\ref{VVV}) are of the order $O(V_\Lambda ^3)$ and closer
inspection of these terms shows that the terms of each type do not cancel
each other. Therefore, the momentum cutoff RG approach in one-loop
approximation does not satisfy the Ward identity (\ref{Ward1}) exactly, but
rather with the accuracy $O(V_\Lambda ^3)$. This accuracy is the same as the
accuracy of the one-loop equations (\ref{OneLoop}). The terms in Eq. (\ref
{VVV}) have a different structure: in the terminology of Ref.\cite{SalmBook}
the terms in the first and second lines of Eq. (\ref{VVV}) correspond to
diagrams with non-overlapping and overlapping loops, respectively. The
presence of the terms with non-overlapping loops prevents the use of more
strict power counting arguments\cite{SalmBook} to find the bounds for the
corresponding contributions responsible for the violation of the Ward
identity.

\begin{figure}[t!]
\psfig{file=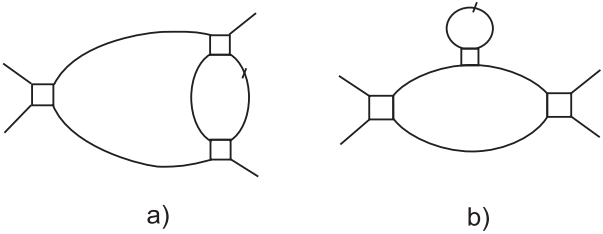,width=80mm,silent=} \vspace{2mm}
\caption{The two-loop corrections to the two-particle vertex: (a) vertex
correction, (b) self-energy correction. The lines with one and two cutting
slashes correspond to the $S^\Lambda$ and $S^{\Lambda ^{\prime }}$
propagators, other lines to $G^{\Lambda ^{\prime }}$ propagators.}
\label{fig:Fig1}
\end{figure}

Consider now the two-loop 1PI RG approach. The two-loop approximation
corresponds to the inclusion of the contribution of the three-particle
vertex $V_\Lambda ^{(6)}$ up to the order $O(V_\Lambda ^3).$ The
corresponding equations read 
\begin{mathletters}
\label{TwoLoop}
\begin{eqnarray}
\frac{d\Sigma ^\Lambda }{d\Lambda } &=&V^\Lambda \circ S^\Lambda
\label{TwoLoopA} \\
\frac{dV^\Lambda }{d\Lambda } &=&V^\Lambda \circ (G^\Lambda \circ S^\Lambda
+S^\Lambda \circ G^\Lambda )\circ V^\Lambda  \label{TwoLoopB} \\
&&+S^\Lambda \circ \int_\Lambda ^{\Lambda _0}d\Lambda ^{\prime }(V^{\Lambda
^{\prime }}\circ G^{\Lambda ^{\prime }}\circ V^{\Lambda ^{\prime }}\circ
G^{\Lambda ^{\prime }}\circ V^{\Lambda ^{\prime }}\circ S^{\Lambda ^{\prime
}})  \nonumber \\
&&+S_\Lambda \circ \int_\Lambda ^{\Lambda _0}d\Lambda ^{\prime }(V^{\Lambda
^{\prime }}\circ (G^{\Lambda ^{\prime }})^2\circ V^{\Lambda ^{\prime }}\circ
S^{\Lambda ^{\prime }}\circ V^{\Lambda ^{\prime }})  \nonumber
\end{eqnarray}
where the last two lines of Eq. (\ref{TwoLoopB}) correspond to the vertex
(Fig. 1a) and self-energy (Fig. 1b) corrections to the one-loop equations.
Performing the same steps as for the one-loop approximation, we obtain a
necessary condition to fulfill Ward identities: 
\end{mathletters}
\begin{eqnarray}
\frac{\partial V^\Lambda }{\partial k_0}=V^\Lambda \circ \frac \partial {%
\partial k_0}(G^\Lambda \circ G^\Lambda )\circ V^\Lambda
\;\;\;\;\;\;\;\;\;\;\;\;\;\;\;\;\;\;\;\;\;\;\;  \label{ddv}
\end{eqnarray}
\begin{eqnarray*}
&&\ \,\,\,\;\,\;\,\;+(G^\Lambda )^2\circ \int_\Lambda ^{\Lambda _0}d\Lambda
^{\prime }(V^{\Lambda ^{\prime }}\circ G^{\Lambda ^{\prime }}\circ
V^{\Lambda ^{\prime }}\circ G^{\Lambda ^{\prime }}\circ V^{\Lambda ^{\prime
}}\circ S^{\Lambda ^{\prime }}) \\
&&\ \;\;\;\;\;+(G^\Lambda )^2\circ \int_\Lambda ^{\Lambda _0}d\Lambda
^{\prime }(V^{\Lambda ^{\prime }}\circ (G^{\Lambda ^{\prime }})^2\circ
V^{\Lambda ^{\prime }}\circ S^{\Lambda ^{\prime }}\circ V^{\Lambda ^{\prime
}})
\end{eqnarray*}
Comparing again this equation with the RG equation for the vertex, Eq. (\ref
{TwoLoop1B}), one can observe that the equations are compatible with the
accuracy $O(V_\Lambda ^4).$ As well as in the one-loop approximation, both
types of terms, with overlapping and non-overlapping loops of order $%
O(V_\Lambda ^4),$ violate the Ward identity (\ref{Ward1}).

However the modification of Eqs. (\ref{TwoLoop}) with the last term replaced
by its local-in-$\Lambda $ analog, i.e. 
\begin{eqnarray*}
&&S^\Lambda \circ \int_\Lambda ^{\Lambda _0}d\Lambda ^{\prime }(V^{\Lambda
^{\prime }}\circ G^{\Lambda ^{\prime }}\circ V^{\Lambda ^{\prime }}\circ
G^{\Lambda ^{\prime }}\circ V^{\Lambda ^{\prime }}\circ S^{\Lambda ^{\prime
}}) 
\begin{array}{c}
\rightarrow
\end{array}
\\
&&\ \ \ \ \ \ \ \ \ \ \ \ \ S^\Lambda \circ V^\Lambda \circ G^\Lambda \circ
V^\Lambda \circ G^\Lambda \circ V^\Lambda \circ G^\Lambda
\end{eqnarray*}
leads to improvement in the fullfilment of the Ward identities. The
corresponding equations read 
\begin{mathletters}
\label{TwoLoop1}
\begin{eqnarray}
\frac{d\Sigma ^\Lambda }{d\Lambda } &=&V^\Lambda \circ S^\Lambda
\label{TwoLoop1A} \\
\frac{dV_\Lambda }{d\Lambda } &=&V^\Lambda \circ \frac d{d\Lambda }%
(G^\Lambda \circ G^\Lambda )\circ V^\Lambda  \label{TwoLoop1B} \\
&&+S^\Lambda \circ \int_\Lambda ^{\Lambda _0}d\Lambda ^{\prime }(V^{\Lambda
^{\prime }}\circ G^{\Lambda ^{\prime }}\circ V^{\Lambda ^{\prime }}\circ
G^{\Lambda ^{\prime }}\circ V^{\Lambda ^{\prime }}\circ S^{\Lambda ^{\prime
}})  \nonumber
\end{eqnarray}
The difference between (\ref{TwoLoop}) and (\ref{TwoLoop1}) is itself of the
order of $O(V^4),$ i.e. of the order of the terms neglected in the two-loop
approximation. The advantage of the modified two-loop RG equations, Eq. (\ref
{TwoLoop1}) is that they violate the Ward identity (\ref{Ward1}) by terms
with overlapping loops only. Since such terms can correct only vertices but
not the internal Green functions, one can consider Eqs. (\ref{TwoLoop1}) as
one-particle self-consistent. The absence of the terms with non-overlapping
loops in the difference between the r.h.s. and the l.h.s. of the Ward
identity (\ref{Ward1}) in this case is a consequence of the non-trivial
cancellation between derivatives of the self-energy contained in the
propagators of 1PI scheme (\ref{GMC}) and the vertex corrections.

In particular, in the ladder approximation, when one selects only diagrams
in the appropriate particle-hole channel and neglects the last term in the
r.h.s. of Eq. (\ref{TwoLoop1B}), Eqs. (\ref{TwoLoop1}) take the form 
\end{mathletters}
\begin{mathletters}
\label{TwoLoopLadd}
\begin{eqnarray}
\frac{d\Sigma ^\Lambda }{d\Lambda } &=&V^\Lambda \circ S^\Lambda
\label{TwoLoopLadd1} \\
\frac{dV_\Lambda }{d\Lambda } &=&[V^\Lambda \circ \frac d{d\Lambda }%
(G^\Lambda \circ G^\Lambda )\circ V^\Lambda ]_{\text{ph,\thinspace ladder}}
\label{TwoLoopLadd2}
\end{eqnarray}
These ladder equations differ from those in the one-loop approximation (\ref
{OneLoop}) only by the replacement $S_\Lambda \rightarrow dG_\Lambda
/d\Lambda $ in the equation for the two-particle vertex and fulfill Ward
identities exactly at any scale $\Lambda $. The solution of these equations
leads to RPA-type vertices $V_\Lambda ,$ which satisfy standard integral
equations 
\end{mathletters}
\begin{eqnarray}
V_{kp;pk}^\Lambda &=&V_{kp;pk}^0+\sum_{k^{\prime }}V_{kk^{\prime };k^{\prime
}k}^0(G_{k^{\prime }}^\Lambda )^2V_{k^{\prime }p;pk^{\prime }}^\Lambda 
\nonumber \\
V_{kp;kp}^\Lambda &=&V_{kp;kp}^0+\sum_{k^{\prime }}V_{kk^{\prime
};kk^{\prime }}^0(G_{k^{\prime }}^\Lambda )^2V_{k^{\prime }p;pk^{\prime
}}^\Lambda  \nonumber \\
&&\ +\sum_{k^{\prime }}(V_{kk^{\prime };k^{\prime }k}^0-2V_{kk^{\prime
};kk^{\prime }}^0)(G_{k^{\prime }}^\Lambda )^2V_{k^{\prime }p;k^{\prime
}p}^\Lambda  \label{VV1}
\end{eqnarray}
and the mean-field-type result for the self-energy, 
\begin{equation}
\Sigma _k^\Lambda =\sum_p(V_{kp;pk}^0-2V_{kp;kp}^0)G_p^\Lambda  \label{SS}
\end{equation}
with the full (dressed) Green function $G^\Lambda .$ In the limit $\Lambda
\rightarrow 0$ Eqs. (\ref{VV1}) and (\ref{SS}) reduce to standard RPA and
mean-field results, respectively.

For the non-ladder case and the regular curved Fermi surface one can use
power counting arguments \cite{SalmBook,SalmHonR} to show that the equations
(\ref{TwoLoop1}) violate the Ward identity (\ref{Ward1}) by finite
(non-divergent in the $\Lambda \rightarrow 0$ limit) terms $O(V_\Lambda ^4)$%
. One can expect that these equations will better satisfy Ward identities
for both regular and singular Fermi surfaces due to the smallness of the
contributions of the terms with overlapping loops\cite{SalmBook}. Note that
the replacement of the last term in Eq. (\ref{TwoLoop1}) by its local analog
and the additional replacement $S_\Lambda \rightarrow dG_\Lambda /d\Lambda $
in the same term (which again produces finite errors of the order $%
O(V_\Lambda ^4)$) can further improve the fulfilllment of Ward identities.
The analysis of this possibility requires however the consideration of the
three-loop terms, which is beyond the scope of the present paper.

Therefore, in general, the momentum-cutoff RG scheme (\ref{GMC}) at two-loop
order satisfies the frequency component of the Ward identity (\ref{Ward})
with the accuracy $O(V_\Lambda ^4)$. The same situation holds for the
momentum component of the Ward identity (\ref{Ward}) in the frequency cutoff
RG approach, Eq. (\ref{GFC}). The frequency component of the Ward identity (%
\ref{Ward}) in frequency cutoff scheme and momentum component in
momentum-cutoff scheme are violated because of additional terms coming from
the differentiation of the cutoff functions\cite{Enss}. At the same time, in
the temperature cutoff RG scheme (\ref{GTC}), which does not contain any
cutoff in momentum or frequency space, similar calculations show that both,
frequency- and momentum-components of the Ward identity are satisfied at
two-loop order up to the same accuracy $O(V_\Lambda ^4)$.

In conclusion, we have considered the fulfillment of Ward identities in the
one- and two-loop RG approach. The frequency components of Ward identities
in the momentum-cutoff scheme and the momentum components in the
frequency-cutoff scheme are generally satisfied up to terms of $O(V_\Lambda
^3)$ in the one-loop approximation and $O(V_\Lambda ^4)$ in the two-loop
approximation. The one-particle self-consistent version of the two-loop RG
equations (\ref{TwoLoop1}) leads to smaller errors in the fulfillment of
Ward identities due to absence of terms with non-overlapping loops. In the
ladder approximation this version of the two-loop equations satisfies Ward
identities exactly.

I am grateful to W. Metzner, M. Salmhofer, A. Kampf, T. Enss, and A. O.
Anokhin for stimulating discussions.

\end{document}